\documentclass{aa}  
\usepackage{graphicx}
\usepackage{graphicx}
\usepackage{savesym}
\usepackage{amsmath}
\savesymbol{iint}
\usepackage{txfonts}
\restoresymbol{TXF}{iint}
\usepackage{amssymb}
\usepackage{epsfig}
\usepackage{natbib}
\usepackage{multirow}

\begin{document}

   \title{Evolution of tidal disruption candidates \\
	  discovered by \emph{XMM--Newton} }
   
   \author{P. Esquej
          \inst{1}\fnmsep\inst{3},
	  R.D. Saxton\inst{2},
	  S. Komossa\inst{1},
	  A.M. Read\inst{3},
          M.J. Freyberg\inst{1},
	  G. Hasinger\inst{1}, \\
	  D.A. Garc\'{i}a-Hern\'{a}ndez\inst{5}, 
	  H. Lu\inst{4},
          J. Rodr\'{i}guez Zaur\'{\i}n\inst{6},
	  M. S\'{a}nchez-Portal\inst{2}
	  \and
	  H. Zhou\inst{1}
	  }

   \offprints{P. Esquej \\ \email{pili@mpe.mpg.de}}

   \institute{
	Max-Planck-Institut f\"{u}r extraterrestrische Physik, Giessenbachstrasse 1, 85748 Garching, Germany
	\and
	ESA/ESAC, Apartado 78, 28691 Villanueva de la Ca\~{n}ada, Madrid, Espa\~{n}a 
	\and
	Dept. of Physics and Astronomy, Leicester University, Leicester LE1 7RH, U.K.
	\and
	Center for Astrophysics, University of Science and Technology of China, Hefei, Anhui, 230026, P.R. China
	\and
	Instituto de Astrof\'{i}sica de Canarias, La Laguna E-38200, Tenerife, Spain
	\and
	Dept. of Physics and Astronomy, University of Sheffield, Sheffield S3 7RH, U.K.\\
             }

   \date{Received May 2, 2008; accepted July 22, 2008}
   

  \abstract
   {It has been demonstrated that active galactic nuclei are powered by gas accretion onto supermassive black holes located at their centres. The paradigm that the nuclei of inactive galaxies are also occupied by black holes was predicted long ago by theory. In the last decade, this conjecture was confirmed by the discovery of giant-amplitude, non-recurrent X-ray flares from such inactive galaxies and explained in terms of outburst radiation from stars tidally disrupted by a dormant supermassive black hole at the nuclei of those galaxies. }
   {Due to the scarcity of detected tidal disruption events, the confirmation and follow-up of each new candidate is needed to strengthen the theory through observational data, as well as to shed new light on the characteristics of this type of events. }
   {Two tidal disruption candidates have been detected with \emph{XMM--Newton} during slew observations. Optical and X-ray follow-up, post-outburst observations were performed on these highly variable objects in order to further study their classification and temporal evolution.}
   {We show that the detected low-state X-ray emission for these two candidates has properties such that it must still be related to the flare. The X-ray luminosity of the objects decreases according to theoretical predictions for tidal disruption events. At present, optical spectra of the sources do not present any evident signature of the disruption event. In addition, the tidal disruption rate as derived from the \emph{XMM--Newton} slew survey has been computed and agrees with previous studies.}
   {}

   
   \authorrunning{P. Esquej et al.}
   \titlerunning{Evolution of tidal disruption candidates
   }
   \maketitle


\section{Introduction}\label{sec:intro}
According to our current picture of galaxy demography, massive dark objects could reside at the centres of many galaxy bulges \citep{Kormendy, Magorrian}. While it has been demonstrated that active nuclei powered by gas accretion onto supermassive black holes (SMBHs) populate the cores of a number of galaxies, confirmation of their dormant state lurking in non-active galaxies is difficult to obtain. 

Chronologically, quasars were more abundant in the early Universe at \emph{z}$\sim$2 than at present so dead quasar engines are expected to be enclosed in the nuclei of otherwise inactive galaxies. Evidence connecting the genesis of SMBHs and the possible hierarchical quasar to galaxy evolution remains elusive. At the lower end of the activity scale and representative of the most common activity level in the local Universe reside low-ionization nuclear emission-line regions (LINERs) \citep{Heckman1980}. Their optical spectra are similar to those of giant elliptical galaxies \citep{Denicolo}, dominated by low-ionization emission lines, that are also similar to the narrow lines in Seyfert galaxies but with different line ratios \citep{Ho2008}. There is much debate concerning LINERs and their powering mechanism caused either by an active nucleus or starburst. Given their activity level and spectral properties they could represent the connection between active and inactive galaxies \citep{Gonzalez2006}.

The paradigm that the cores of inactive galaxies are occupied by quiescent SMBHs was predicted long ago. The discovery of dynamical evidence of a mass concentration with properties of this type of object at the centre of our own galaxy and afterwards in a handful of other galaxies like M31 and NGC~4258 favours this conjecture \citep{Kormendy}. An unavoidable consequence of the existence of remnant SMBHs at the nuclei of optically inactive galaxies is the detection of flare radiation produced when a star is tidally disrupted and accreted by SMBHs \citep{Lidskii, Carter, Rees}. These occurrences, so-called \textit{tidal disruption events}, would generate outbursts of EUV/X-ray radiation decaying on a time scale of several months to years. Although active galactic nuclei (AGN) can also be susceptible to these phenomena, the most unambiguous cases for a stellar disruption come from host galaxies with no - or only faint - permanent activity.

It is expected that tidal flares emit a luminosity peaking in the EUV/X-ray regime. When assuming a black body radiating at the Eddington luminosity, the temperature of the emitting region calculated at the last stable orbit (3${r}_{\rm s}$, where ${r}_{\rm s}$ is the Schwarzschild radius) will be $\sim$60~eV for a 10$^{6}$ $M_{\odot}$ black hole. At low redshifts, the optical and UV bands will be sensitive to the Rayleigh-Jeans tail of the black body radiation, but these events will be best detected in X-rays, with giant amplitudes of variability that will reach factors of hundreds up to several thousand \citep[e.g.][]{Kom04, Halpern2004}.

The comparison of sources from the \emph{ROSAT} PSPC All-Sky Survey (RASS) \citep{Voges} with ROSAT PSPC pointed observations allowed the detection of five tidal disruption candidates. These sources showed extreme X-ray variability and properties in agreement with those expected for tidal disruption events \citep[see][for a review]{Kom02}: extreme X-ray softness in outburst, X-ray peak luminosity up to $\sim10^{45}$ erg~s$^{-1}$ that declines at a rate $t^{-5/3}$ on a timescale from months to years  \citep{Bade, KomBade, KomGreiner, Grupe}. A few candidates have emerged in the UV/optical \citep[e.g.][]{Renzini, Stern, Gezari2008, Kom08}, three of which were subsequently detected in soft X-rays between 1 and 3 years after the high-state detection.

These occurrences are very rare in the sky. Theoretical models postulate that a tidal disruption event should occur once every $10^{3}$ -- $10^{6}$ yr per galaxy \citep{Wang} depending on the stellar density in the nuclear cusp and the SMBH mass. The derived tidal disruption rate from the RASS, although including large uncertainty, lies in agreement with these predictions \citep{Sembay, Donley}. Although previously detected tidal disruption events support theoretical predictions, due to the paucity of candidates  the detection and confirmation of each new case will shed new light on the phenomena here debated.

The identification of tidal disruption events discussed in the present paper presupposes the existence of two large area X-ray sky surveys performed at different epochs. Correlating the \emph{XMM--Newton} Slew Survey Source Catalogue (XMMSL1) \citep{Saxton} with the RASS we found two galaxies which were candidates for tidal disruption events \citep{Esquej}. Here we present follow-up multiwavelength observations performed to investigate the classification of these sources and study their temporal evolution.

This paper is structured as follows: in section (\ref{sec:theory}) we introduce the basics of tidal disruption phenomena, then we define the criteria followed for selecting our sample and present the X-ray and optical properties of the sources. Detailed analysis of follow-up observations is described in section (\ref{sec:obs}). X-ray light curves and estimations of the black hole masses are calculated in the subsequent sections. The tidal disruption rate as derived from the \emph{XMM--Newton} slew survey is derived in section (\ref{sec:tidal_rate}) and our conclusions are finally summarised.

Throughout this paper we will assume a $\Lambda$CDM cosmology with ($\Omega_{\rm M}$,~$\Omega_{\Lambda}$)~=~(0.3,~0.7) and ${H}_{0}$~=~70~ ${\rm km}~{\rm s}^{-1}~{\rm Mpc}^{-1}$. All X-ray fluxes and luminosities here quoted have been corrected for Galactic absorption.

\section{Theory of tidal disruption events}\label{sec:theory}
A star of mass $m_{\star}$ and radius $R_{\star}$ orbiting around a SMBH of mass $M_{\rm BH}$ will be disrupted when it approaches the black hole tidal radius

\begin{equation}
R_{\rm T} = \mu R_{\star}\left( \frac{M_{\mathrm{BH}}} {m_{\star}} \right)^{1/3}, 
\label{eq:tidal_radius}
\end{equation}

\noindent
as this is the distance where the surface gravity of the star equals the tidal force from the hole ($\mu$~is a dimensionless coefficient of order unity). The stellar disruption process is predicted to happen up to black hole masses of 10$^{8}$ $M_{\odot}$. For objects above this limit, the tidal radius is inside the Schwarzchild radius and, therefore, a solar-type star will fall into the black hole without disruption. Atmosphere stripping of giants is still possible for larger black hole masses. 

During its infall towards the central black hole the star experiences a gravitational torque which spins it up and creates a large velocity dispersion within the disturbed material. Once the star is disrupted, half of its debris will have enough energy to exceed the escape velocity and will be consequently ejected on hyperbolic orbits. The remaining half will be bound, returning to pericentre and circularizing \citep{Ulmer, Rees}. This returning material that initially follows highly elliptical orbits, will form an orbiting torus around the black hole due to dissipation and internal shocks. The observable flare of radiation will begin at a time coincident with the accretion of the most tightly bound material as it first returns to pericentre. After one post-disruption orbit, a fraction of the remaining bound debris will be accreted by the hole at about the Eddington rate. \citet{Ayal} showed that as little as 10\% of the total mass of the star would be accreted, unlike previous estimates that claimed that all the bound material (50\% of the star) would be accreted by the black hole. During this stage the soft X-ray luminosity of the galaxy is increased by up to several orders of magnitude and then declines at a rate $t^{-5/3}$ on a timescale from months to years \citep{Evans, Ulmer}.

The tidal disruption rate has been predicted by several authors based on the standard model of collisional loss cone repopulation in a spherical nucleus \citep{Magorrian, Syer, Wang}. In general, this quantity depends on the mass of the black hole, the stellar density, velocities in the galactic nucleus, and the rate at which radial loss cone orbits are depleted/replenished. Wang \& Merrit (2004) derived a tidal rate of $10^{-4}$ $-$ $10^{-5}$ ${\rm yr}^{-1}$ ${\rm Mpc}^{-3}$ for a $10^{6}$ $M_{\odot}$ black hole.

\section{Candidates selection}\label{sec:sample}
The state-of-the-art satellite \emph{XMM--Newton} \citep{Jansen} features high spectral resolution and excellent sensitivity due to the great collecting area of its mirrors coupled with the high quantum efficiency of the EPIC detectors. X-ray objects detected with the EPIC-pn camera on board XMM-Newton while the satellite is maneuvering between observation targets comprise the XMMSL1 source catalogue. 

Our sample selection criteria aimed at finding tidal disruption candidates through comparison of the XMMSL1 with the RASS \citep[see][for details]{Esquej}. For the two resulting objects, NGC~3599 and SDSS~J132341.97+482701.3, no prior AGN-like activity had been recorded from optical observations according to previous published measurements (but see section \ref{subsec:NGC3599}). Both sources showed high variability and a soft X-ray spectrum as regards the slew observations. All these properties favoured the interpretation of the flaring events as tidal disruption candidates.

\begin{figure*}[ht]
   \centering
   {\rotatebox[]{0}{\includegraphics[width=17cm, height=8.5cm]{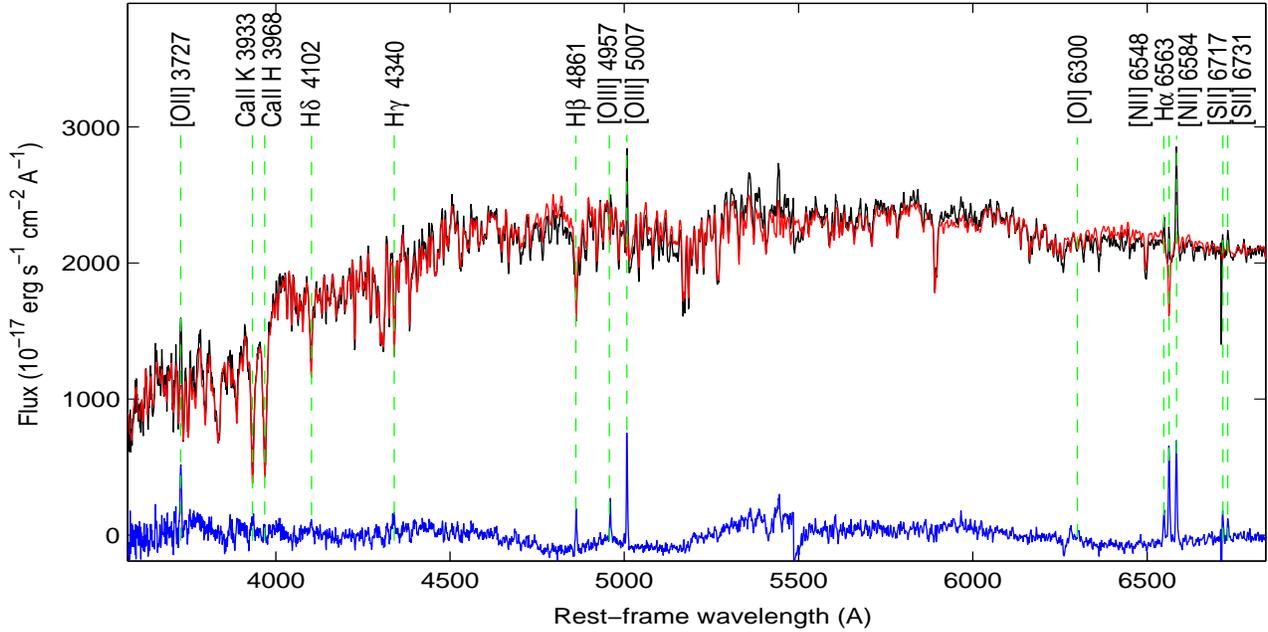}}}
  
     \caption{Detailed fits to the optical spectrum of NGC~3599. Flux calibrated data from \cite{Caldwell} are presented in black. The best fitting stellar population as derived from the EL-ICA approach is shown in red. The subtracted spectrum in blue illustrate the galactic nuclear region. Line identifications are given in the figure.}               
        \label{fig:NGC3599_Caldwell}
\end{figure*}

\subsection{NGC 3599}\label{subsec:NGC3599}
The slew source XMMSL1~J111527.3+180638 was detected in 2003$-$11$-$22 with an EPIC-pn 0.2-12 keV count rate of 5.5 cts s$^{-1}$. Through catalogue cross-correlation it was found that its counterpart is the SA0 galaxy NGC~3599 at redshift 0.0028 \citep{Denicolo}. Within the early-type galaxies, NGC~3599 is a power-law galaxy (its brightness profile has a power-law shape) with a compact nuclear component not associated with non-thermal activity (like those found at the centres of M31 and M33) \citep{Lauer}. Although no public optical spectroscopic observations of this source were available, a detailed literature search revealed that NGC~3599 was already briefly included in other samples with different aims. Flux calibrated optical data were kindly provided to us from the sample of \citet{Caldwell} and then used to characterise the source in the pre-outburst stage. 

The long-slit optical spectrum of NGC~3599 was taken in 2000$-$02$-$05 at the F. L. Whipple Observatory with the FAST spectrograph located in the 1.5m Tillinghast telescope \citep[see][for observational details and data reduction procedure]{Caldwell}. Prominent absorption features in combination with emission lines appear in the integrated spectrum. In order to further study the emission-line spectrum we first subtracted the starlight component. This step in the data analysis reduction is crucial to derive an accurate nuclear source classification based on emission lines ratios. Two independent approaches to the stellar population reconstruction have been implemented both using high resolution synthetic spectra from the evolutionary synthesis models of \citet{Bruzual} and assuming a Salpeter initial mass function \citep{Salpeter}. These techniques comprise the Ensemble Learning Independent Component Analysis (EL-ICA) and the CONFIT codes. Discussion about technical details on both resources is beyond the scope of the current paper and presented elsewhere in the literature (see references below).

The EL-ICA code presupposes that the stellar contribution is a linear combination of six non-negative independent components (ICs) that arise from the compaction of the 1326 simple stellar populations (SSPs) of \citet{Bruzual} \citep[][and references therein]{Lu}. Through minimization of the $\chi_{red}^2$ parameter, the optimum fit to our data is obtained with a single stellar population of 2.1~Gy and solar metallicity. Fig.~\ref{fig:NGC3599_Caldwell} shows the data (black), stellar modelling (red) and the subtracted spectrum that accounts for the galactic nuclear emission (blue). 

Conversely, the CONFIT approach consists of modelling the continuum spectral shape using two stellar components allowing them to have different reddening and applying $\chi^2$ statistics to derive the optimum fit \citep{Robinson, Rodriguez}. The continuum flux is measured in a number of wavelength bins (54 in this case) distributed over the spectral regions free of emission lines and atmospheric absorption features. By computing the contribution of the different stellar components to reproduce the complete spectrum, the best fit resulted from a single stellar population of 2~Gy, which is fairly consistent with the outcome of the EL-ICA approach. . 

Spectral emission lines have been fitted using a single-component Gaussian. NGC~3599 exhibits narrow emission lines of about 100~km~s$^{-1}$ width. Intensity ratios of [O~III]~5007~\AA/H$\beta$=3.5($\pm$ 1.0), [N~II]~6584~\AA/H$\alpha$=1.1($\pm$ 0.2) were measured from the analysis of the decomposed spectrum as returned by the EL-ICA code. As a comparison, those calculated from the CONFIT procedure provide [O~III]5007~\AA/H$\beta$=3.0($\pm$ 0.2) and [N~II]6584~\AA/H$\alpha$=1.1($\pm$ 0.3). Both methods place the source in the region between Seyfert and LINER galaxies according to emission line diagnostics \citep{Kauffmann}. Due to large uncertainties in the [SII] lines fluxes, we can not use such measurements for a reliable input in the diagnostics for the source classification.

In the X-ray regime, NGC~3599 was detected in a high state by \emph{XMM--Newton} with respect to the RASS having a 0.2--2 keV EPIC-pn vs PSPC-$2\sigma$ upper limit flux ratio of 88. In addition, the source also lies in the field of view of two \emph{ROSAT} PSPC pointed observations and corresponding upper limits imply flux ratios of 297 and 100 respectively. Both \emph{XMM} and ROSAT 0.2--2 keV fluxes were estimated from count rates using PIMMS assuming a typical near-maximum tidal event spectrum, i.e. a black body model with \emph{kT}=0.07 keV modified by Galactic absorption (Komossa 2002). The unabsorbed X-ray luminosity in the soft 0.2--2 keV energy band at the time of the slew observation was calculated to be $5.1(\pm 0.7)\times10^{41}$ erg~s$^{-1}$. As the source lies within the Virgocentric flow the luminosity distance is not a relevant parameter for the calculation of the X-ray luminosity, instead, a distance of 23 Mpc \citep{Lauer2007} has been used. 

We have analysed the slew observation corresponding to the detection of NGC~3599. The object does not suffer from pile-up problems so the pn spectrum was extracted with patterns 0--4. The source spectrum has been extracted from a circle of radius 60 arcsec centred on the object.  An adjacent region of the same radius has been used for the background. As software to produce response matrices for slew sources is not available, we have created the ancillary file using an \textit{ad hoc} method of averaging the effective area along the track of the slew \citep[see][for a full explanation of the technique]{Read2008}. For the redistribution matrix we used the canned response file \textit{epn\_ff20\_sdY6\_v6.9.rmf}, appropriate for the average position of the source in the detector. Due to the low signal-to-noise of the object, a binning of 5 counts per spectral channel has been used for the spectral fitting. This is enough to perform a quantitative analysis although not statistically significant. Therefore, we can only attempt to derive a rough spectral shape of the object applying simple spectral models (Table~\ref{tab:spectral_parameters_slew}).

\begin{figure}
\centering

   {\rotatebox[]{270}{\includegraphics[width=6cm]{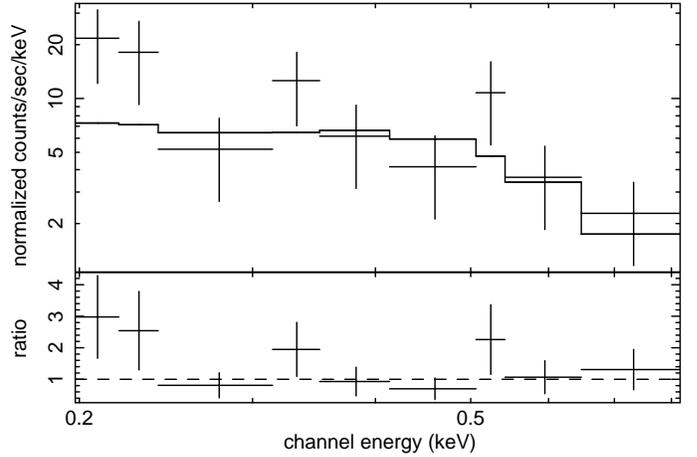}}}
 \vspace{-1.3cm}    
\caption{\emph{XMM--Newton} EPIC-pn slew X-ray spectra of NGC~3599 fitted by the black body model plus Galactic absorption. Data/model residuals are presented at the bottom.}
\label{fig:NGC_slew_spectrum}
\end{figure}

As a result from the spectral analysis using XSPEC 11.3.2, a good fit to the data ($\chi^2_{\rm red}$=1.08) is provided by a black body model of \emph{kT}=95$^{+4}_{-3}$ eV modified by Galactic absorption (Fig.~\ref{fig:NGC_slew_spectrum}). This temperature is consistent (within the errors) with measurements from previous candidates although at the upper range of temperatures \citep{Kom02}. The corresponding 0.2--2 keV X-ray luminosity from the XSPEC fit is $5.5\times10^{41}$ erg~s$^{-1}$. Using a power-law model, an equally good fit is achieved with $\Gamma_{\rm X}$=3.3$^{+1.25}_{-1.12}$. A range of spectral models for the high-state spectrum have been used in PIMMS to estimate the flux at the time of the slew observation. For black body models with temperature within 0.04--0.1 keV and power-law models with X-ray slopes of 2.9--3.5, the luminosities lie in the range $4.4\times10^{41}$--$6.5\times10^{41}$ erg~s$^{-1}$. The median luminosity from the different spectral models is obtained with the black body model at 0.07~keV, i.e. the typical temperature for a tidal disruption event. The extrapolation of the slew count rate in the 0.2--12~keV to calculate an upper limit for the 2--12~keV luminosity gives $2\times10^{40}$ erg~s$^{-1}$ for the power-law model with $\Gamma_{X}$=3.3 (and $7\times10^{35}$ erg~s$^{-1}$ for the black body model with \emph{kT}=0.095~keV).

\begin{table}
  
\caption{Spectral fit parameters (including 90\% confident errors) of the EPIC-pn slew spectrum of NGC~3599.}             
\label{tab:spectral_parameters_slew}      
\centering

\sbox{\strutbox}{\rule{0pt}{0pt}}
\begin{tabular}{c|c|c|c}        

\hline
& & &\\[\smallskipamount]
\textbf{Model} & \textbf{$\Gamma_{\rm X}$} & \textbf{\emph{kT} (keV)} & \textbf{$\chi^{2}$(d.o.f.)}  \\ 

\hline
                  
& & &\\[\smallskipamount]

	  wabs(bb)          & (...)	& 0.095$^{+0.004}_{-0.003}$	&7.6(7)\\
& & &\\[\smallskipamount] 
	   wabs(po)         & 3.33$^{+1.25}_{-1.12}$ &  (...)   & 6.1(7)  \\ 
& & &\\[\smallskipamount]
	   wabs(po+bb)      & 2.86$^{+1.30}_{-1.31}$ & 0.009$^{+0.003}_{-0.001}$ & 4.2(5)  \\  
& & &\\[\smallskipamount]
  wabs(po+MEKAL)   & 4.45$^{+0.40}_{-0.64}$ & 0.21$^{+79}_{-0.13}$ & 5.6(5) \\
& & &\\[\smallskipamount]
\hline

\end{tabular}
\end{table}

\begin{figure*}[ht]

\centering
{\rotatebox[]{0}{\includegraphics[width=17cm, height=8.5cm]{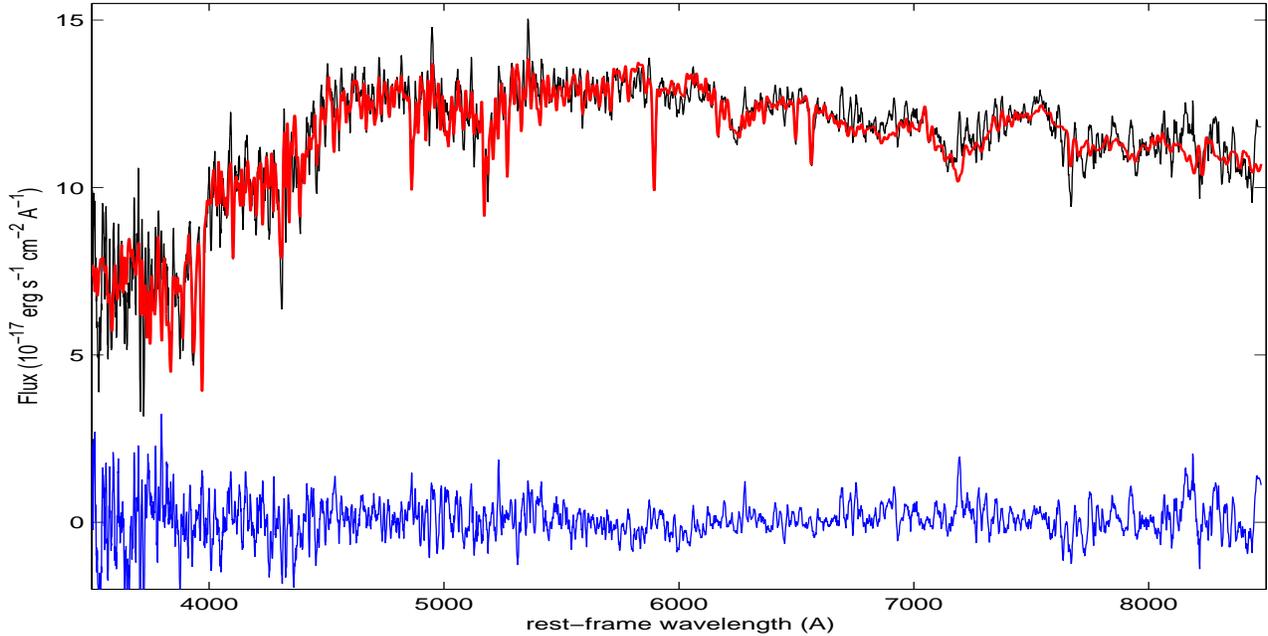}}}

\caption{Detailed fits to the optical spectrum of SDSS~J132341.97+482701.3. Flux calibrated data from the SDSS server are presented in black. The best modelling to the continuum as derived from the EL-ICA approach is shown in red and corresponds to a stellar population of 6.25 Gy and metallicity \textit{Z}=0.4 $Z_{\odot}$. The residual spectrum resulted from the subtraction of the best fit model to the observed spectrum is shown in blue.}

\label{fig:J1323_SDSS}
\end{figure*}
\subsection{SDSS~J132341.97+482701.3}\label{subsec:J1323}
XMMSL1~J132342.3+482701 was detected in 2003$-$12$-$01 with a count rate of 2.08 cts s$^{-1}$ in the slew total energy band. Its counterpart, SDSS~J132341.97+482701.3 (named as SDSS~J1323 henceforward), can be classified as a non-active galaxy as its optical spectrum only shows absorption line features from the atmospheres of individual stars. Prior to its detection as tidal disruption candidate, no reference regarding this source is present in the literature. 

Optical data from a pre-outburst observation (performed seven months before the corresponding slew exposure) were obtained from the Sloan Digital Sky Survey\footnote[1]{http://cas.sdss.org/astro/en} (SDSS) and modelled using the EL-ICA method. No emission lines could be detected and the continuum can be best fitted by a single stellar population of 6.25 Gy and a metallicity of \emph{Z}=0.4 $Z_{\odot}$ (Fig. \ref{fig:J1323_SDSS}). The galaxy is at a redshift of \emph{z}=0.0875 and its stellar velocity dispersion has a value of $\sigma$=80($\pm$10) km~s$^{-1}$ as derived from the optical spectrum. As no emission lines that could bias the source classification appear in the spectrum, no additional information can arise from applying the CONFIT approach to the stellar subtraction.

The X-ray slew detection of the object showed a flux variation in the 0.2--2 keV energy band of a factor 83 between the slew detection and its RASS upper limit. The redshift of the source implies a luminosity distance of 399~Mpc, that allowed us to derive a soft unabsorbed X-ray luminosity of $4.8\times10^{43}$ erg~s$^{-1}$ from the slew observation using the black body model for tidal disruption events near outburst. The slew observation contained
too few counts, all of them in the 0.2--2 keV energy band, to allow spectral fitting. 

\section{Follow-up observations}\label{sec:obs}

\subsection{X-ray observations}\label{subsec:X-ray_obs} 

\subsubsection{\emph{XMM--Newton} observations}\label{subsubsec:XMM_obs}
\emph{XMM--Newton} observations of 5 ks of both NGC~3599 and SDSS~J1323 were performed roughly two years after the corresponding slew exposures and then analysed with SAS 7.1.0 (see Table~\ref{tab:observations} for details of the observations).  All EPIC-pn and MOS observations were taken in Full Frame (FF) mode with the medium filter in place. High background periods produced by intense soft proton fluxes were rejected during the data reduction process by defining good time intervals within the complete observation window. Source photons were extracted from circular regions of 30 arcsec radius centred on the object positions. Circular source-free regions on the same chip and radius of 60 arcsec have been used to characterise the background. The targets do not suffer from pile-up problems, so pn spectra were extracted with patterns 0 to 4 and MOS spectra from 0 to 12. Ancillary files and response matrices were generated by the \textit{arfgen} and \textit{rmfgen} SAS tasks respectively in order to convert the counts to physical units during the spectral analysis. All EPIC spectra were binned to oversample the instrumental resolution and to have no less than 20 counts in each background-subtracted spectral channel. Due to the low signal-to-noise ratio of the MOS spectra in both objects, the grouping resulted in at most three spectral points from each MOS observation. As a consequence, we only considered pn data for the source modelling performed using XSPEC. 

\begin{table}
  
\caption{Summary of the X-ray follow-up observations.}             
\label{tab:observations}
\centering
 \sbox{\strutbox}{\rule{0pt}{0pt}}
\begin{tabular}{c c c c}

\hline
& & &\\[\smallskipamount]
\textbf{Source name} & \textbf{Mission/Instrument} &  \textbf{ObsID} & 
\textbf{Exposure} \\    
 & &  &\\[\smallskipamount]
& &  & \textbf{(ks)} \\
  
& &  &\\[\smallskipamount]
\hline                 

& &  &\\[\smallskipamount] 
\multirow{2}{*}{NGC~3599} & \emph{XMM--Newton} /EPIC & 2006-06-23 &  5 \\
& &  &\\[\smallskipamount]

	& \emph{Swift}/XRT  & 2006-12-01 &  5\\
& &  &\\[\smallskipamount]

\hline
& &  &\\[\smallskipamount]

\multirow{2}{*}{SDSS~J1323} & \emph{XMM--Newton} /EPIC & 2006-07-15 &  5 \\
& &  &\\[\smallskipamount]

              & \emph{Swift}/XRT  & 2007-01-18 &  5\\
& &  &\\[\smallskipamount]
\hline   

\end{tabular}
\end{table}

We initially performed spectral fits on both sources using a simple black body model with cold absorption to draw a comparison with the high-state spectrum of a typical tidal disruption event. In both sources the black-body model resulted in a poor fit. The post-flare spectra appear to have a roughly power-law shape. One-temperature thermal models often used to fit the spectra of normal galaxies (Raymond-Smith and MEKAL; \citep{Fabbiano}) also do not fit well, indicating that we continue to see the remnant of the outburst. From the spectral analysis, no improvement was achieved when including intrinsic absorption, hence all spectral fits were performed assuming Galactic foreground absorption values (${N}_{\rm H}$=$1.42\times10^{20}$ and $1.17\times10^{20}$ cm$^{-2}$ for NGC~3599 and SDSS~J1323 respectively) inferred from HI observations \citep{Dickey}. For both objects the hard X-ray behaviour could not be perfectly constrained by model fitting as a single data bin covering a wide energy range is present in that area of the spectra. Detailed analysis of the individual objects will be described below and spectral parameters have been compiled in Table~\ref{tab:spectral_parameters}.
 
\begin{table*}[ht]
  
\caption{Spectral fit parameters of the EPIC-pn spectra of the sources. 90\% confident errors are quoted when $\chi_{red}^{2}<$2.0}             
\label{tab:spectral_parameters}      
\centering

\sbox{\strutbox}{\rule{0pt}{0pt}}
\begin{tabular}{c|c|c|c|c|c}        

\hline
& & & &\\[\smallskipamount]
\textbf{Source} & \textbf{Model} & \textbf{$\Gamma_{X}$} & \textbf{Parameters} & \textbf{$\chi^{2}$(d.o.f.)} & \textbf{$F_{\rm 0.2-2 keV}$} \\ 

& & & &  & [$10^{-13}$ erg~${\rm s}^{-1}$~${\rm cm}^{-2}$] \\
\hline
                  
& & & &\\[\smallskipamount]

	 &  wabs(po)         & 3.02$^{+0.22}_{-0.21}$ &  (...)    & 20.61(18) & $2.93^{+0.51}_{-0.62}$ \\ 
& & & &\\[\smallskipamount]
	 &  wabs(po+bb)      & 2.81$^{+0.24}_{-0.23}$ & \emph{kT}=22.3$^{+3.9}_{-12.0}$ eV & 15.38(16) & $3.91^{+2.70}_{-1.36}$\\  
& & & &\\[\smallskipamount]
NGC~3599 &  wabs(po+MEKAL)   & 3.77$^{+0.40}_{-0.64}$ & \emph{kT}=1.99$^{+11.98}_{-0.65}$ keV & 14.54(16) & $3.24^{+0.64}_{-1.73}$ \\
& & & &\\[\smallskipamount]
         &  wabs(po+absori)  & 3.02$^{+0.36}_{-0.26}$ & ${N}_{\rm H,abs}$=$0.34^{+0.30}_{-0.26}\times10^{22}$~${\rm cm}^{-2}$; $\xi$=4.32$^{+34.34}_{-3.40}$    & 16.16(16) & 3.07$^{+0.62}_{-2.57}$ \\ 
& & & &\\[\smallskipamount]
         &  wabs(po+zpcfabs) & 3.48$^{+1.30}_{-0.50}$ & \emph{CvFr}=$0.31^{+0.21}_{-0.54}$; ${N}_{\rm H,abs}$=$0.46^{+60.85}_{-0.18}\times10^{22}$~${\rm cm}^{-2}$ & 17.42(16) & 3.29$^{+0.78}_{-2.55}$ \\
& & & &\\[\smallskipamount]
\hline
& & & &\\[\smallskipamount]
 & wabs(po)         & 3.37 & (...) & 12.47(5) & $0.58^{+0.20}_{-0.33}$ \\ 
& & & &\\[\smallskipamount]
			&  wabs(po+bb)      & 5.26$^{+1.53}_{-0.66}$ & \emph{kT}=33.1$^{+66.6}_{-10.6}$ eV & 4.71(3) & $0.98^{+0.31}_{-0.58}$\\  
& & & &\\[\smallskipamount]
SDSS~J1323                     & wabs(po+MEKAL)   & 5.16$^{+1.60}_{-0.35}$ & \emph{kT}=1.72$^{+52.59}_{-0.44}$ keV & 4.21(3) & $0.94^{+0.32}_{-0.56}$ \\
& & & &\\[\smallskipamount]
			&  wabs(po+absori)  & 3.14 & ${N}_{\rm H,abs}$=$0.84\times10^{22}$~${\rm cm}^{-2}$; $\xi$=8.10    & 11.68(3) & 0.63$^{+0.34}_{-0.57}$ \\ 
& & & &\\[\smallskipamount]
			&  wabs(po+zpcfabs) & 4.70$^{+0.58}_{-0.80}$ & \emph{CvFr}=$0.99^{+0.01}_{-0.14}$; ${N}_{\rm H,abs}$=$0.81^{+0.76}_{-0.34}\times10^{22}$~${\rm cm}^{-2}$ & 4.07(3) & 0.90$^{+0.34}_{-0.57}$ \\
& & & &\\[\smallskipamount]
			& wabs(po+edge)         & 3.08$^{+0.47}_{-0.36}$ & \emph{kT}=0.49$^{+0.07}_{-0.06}$ keV & 1.69(3) & $0.73^{+0.29}_{-0.42}$ \\ 
& & & &\\[\smallskipamount]
\hline   
                                
\end{tabular}
\end{table*}

\begin{figure*}[ht]
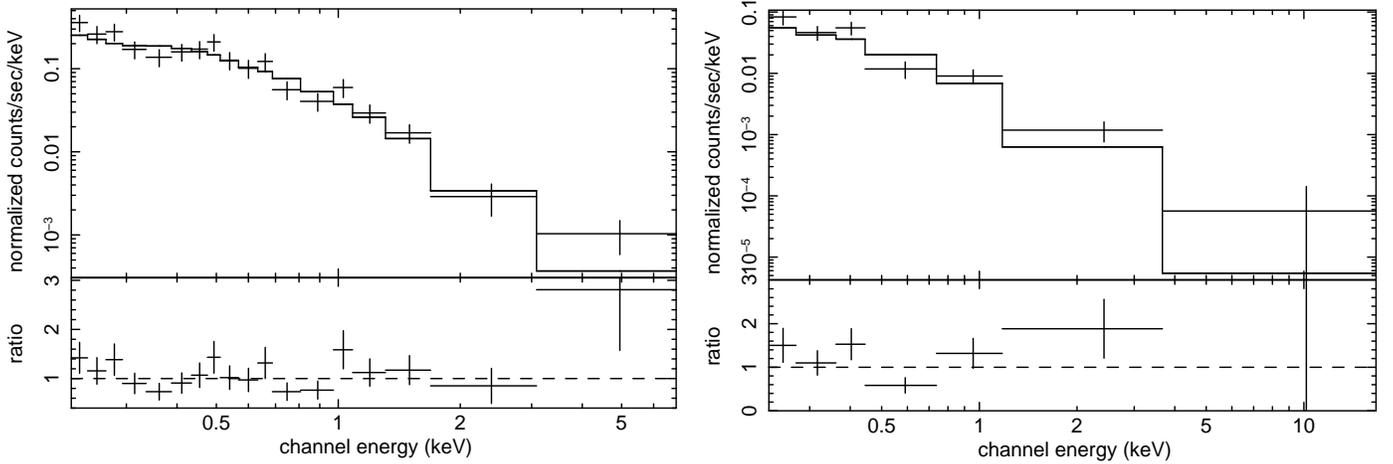


\begin{tabular}{cc}
{\rotatebox[]{270}{\includegraphics[width=6cm]{NGC_XMM_powerlaw.ps}}}
\hspace{0.2cm}
{\rotatebox[]{270}{\includegraphics[width=6.1cm]{SDSS_XMM_powerlaw.ps}}}
\end{tabular}
\vspace{-1.2cm}
     \caption{\emph{XMM--Newton} EPIC-pn X-ray spectra of NGC~3599 (left panel) and SDSS~J132341.97+482701.3 (right panel) fitted by a power-law model plus Galactic absorption. Data/model residuals are presented at the bottom of each spectrum.
               }  
        \label{fig:XMM_spectra}
    \end{figure*}

X-ray emission from NGC~3599 has been detected in both soft 0.2--2 keV and hard 2--12 keV energy bands with corresponding count rates of 0.115($\pm$0.007) and 0.010($\pm$0.003) cts s$^{-1}$. 
In order to give a first estimation of the spectral shape we fitted a power-law model and obtained a reasonable fit ($\chi^2_{\rm red}$=1.15) with slope of $\Gamma_{X}$=$3.02^{+0.22}_{-0.21}$  but showing considerable residuals especially above 3 keV (Fig.~\ref{fig:XMM_spectra}, left panel). Models including ionized absorbers or partial covers (\textit{absori} or \textit{zpcfabs} in XSPEC) as well as those combinations of power-law and black body or thermal component (both using abundances fixed to solar value and treated as free parameter for the MEKAL model) do not provide a significant improvement. In all cases the F-test gives a confidence level of less than 85\% with respect to the single power-law fit. In addition, the reliability of this models is not demonstrated as 90\% confident errors on the parameters can not be constrained within sensible values. In any case the presence of a steep power-law is always required. Some spectral features remained unfitted as shown by the model residuals but with the present data no further constraints can be derived in this respect. This is just an indication that the real model is more complex than the ones here proposed as several emission mechanisms might contribute to the observed X-ray spectra. The unabsorbed luminosity at the time of the \emph{XMM--Newton} observation derived from the simple power-law model is $1.8^{+0.3}_{-0.4}\times10^{40}$ erg~s$^{-1}$, i.e. a factor 27 lower than during the slew exposure. By extrapolating the best fit power-law model to high energies, we obtained a 2--12 keV luminosity of $0.2^{+0.1}_{-0.1}\times10^{40}$ erg~s$^{-1}$.

For SDSS~J1323 the simple power-law model modified by cold absorption yields a reasonable fit in the soft X-rays but leaves an apparent high energy excess (Fig.~\ref{fig:XMM_spectra}, right panel). This model can be best fit using a steep power-law index $\Gamma_{X}$=3.37 and the resulting $\chi^2_{\rm red}$ is 2.49. Confidence errors on the slope could not be derived due to the high value of the $\chi^2_{\rm red}$. The same models as for NGC~3599 have been used in the analysis of SDSS~J1323. Although the derived $\chi^2_{\rm red}$ is closer to unity using some of the complex models, unusually steep slopes are required and fitting parameters can not be constrained. No significant improvement was achieved with respect to the simple power-law model as confirmed by the F-test. Given that the spectrum seems to present an absorption feature around 0.5 keV, we tried to fit it with a power-law plus edge model. A simple edge at an energy of $0.49^{+0.07}_{-0.06}$ keV gives a good fit ($\chi^2_{\rm red}=0.56$) albeit with 
a very small number of degrees of freedom. This is consistent with absorption by cold Oxygen at the redshift of the source. Although the same feature appears when grouping the data with different spectral binnings the low signal-to-noise precludes further interpretation. Count rates in the soft and hard X-ray bands are 0.022($\pm$0.002) and 0.002($\pm$0.001) cts~s$^{-1}$ respectively. The inferred 0.2--2~keV X-ray luminosity as derived from the simple power-law model is $1.1^{+0.4}_{-0.6}\times10^{42}$ erg~s$^{-1}$, implying a decrement of a factor 41 in comparison with the previous slew observation.

\subsubsection{\emph{Swift} observations}\label{subsubsec:Swift_obs}
Follow-up observations of 5 ks have been performed with the X~-ray Telescope (XRT) on board the \emph{Swift} satellite \citep{Gehrels} within the fill-in time program (Table~\ref{tab:observations}). Data from photon counting mode were analysed using XSELECT adopting the standard 0--12 grade filtering. The observing mode that has been used operates in the energy range 0.2--10 keV and retains full imaging and spectroscopic resolution. Target positions have been determined using the \textit{xrtcentroid} tool. Source photons have been extracted from circular regions of 21.2 arcsec based on the count rates of the sources \citep[as inferred from the recommendations in][]{Evans_swift} centred on the objects. For the background determination, annular uncontaminated regions around the sources have been selected containing around a hundred counts in order to minimise the effect of background fluctuations.

\begin{figure*}[ht]
   \centering
   {\rotatebox[]{0}{\includegraphics[width=17cm, height=8.5cm]{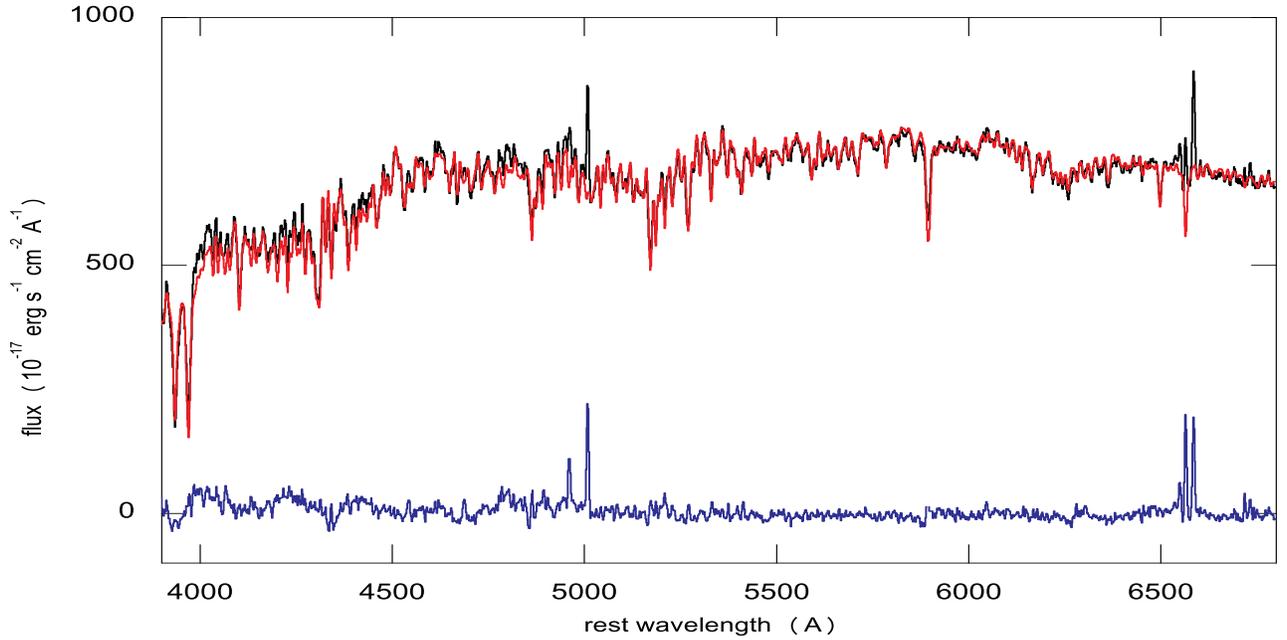}}}
  
     \caption{Detailed fits to the post-outburst optical spectrum of NGC~3599. Flux calibrated data are presented in black and the stellar population is shown in red. The subtracted spectrum in blue illustrate the galactic nuclear region.}               
        \label{fig:NGC3599_INT}
\end{figure*}

Both objects turned out to be extremely faint at the time of the \emph{Swift} observations and, with 23($\pm$5) and 10($\pm$3) counts for NGC~3599 and SDSS~J1323 respectively, we can not apply spectral fits to model the data. In order to estimate the source fluxes we applied a count rate to flux conversion through PIMMS using the same spectral model derived from the previous \emph{XMM} pointed observations (power-law with $\Gamma_{X}$=3.02 and 3.37 for the different objects according to section \ref{subsubsec:XMM_obs}) to our simulations. This is the only reasonable assumption that we can make because the behaviour of the post-outburst spectra is \textit{a priori} unclear. With this hypothesis the source flux of NGC~3599 declined by a further factor of two in comparison with the previous X-ray exposure performed six months earlier. SDSS~J1323 is consistent with constant luminosity within the errors.

\subsection{Optical observations}\label{subsec:optical_obs}
Long-slit optical observations of NGC~3599 and SDSS~J1323 have been performed in order to obtain post-outburst spectral shapes and to search for possible signatures of the tidal disruption events.

Optical spectra have been obtained in December 2007 with the ALFOSC spectrograph on the 2.5 m Nordic Optical Telescope (NOT) in La Palma using the grism 7, that covers a total range from 3850 to 6850~\AA~with an instrumental dispersion of 1.5~\AA~per pixel. Exposure times were 1200 s and 1800 s for NGC~3599 and SDSS~J1323 respectively. The calibration lamp and flat-field exposures were taken before or after any science exposure at the position of the target in order to keep control of possible dispersion changes with the telescope position. Data reduction was performed using the IRAF \footnote{Image Reduction and Analysis Facility (IRAF) software is distributed by the National Optical Astronomy Observatories, which is operated by the Association of Universities for Research in Astronomy, Inc., under cooperative agreement with the National Science Foundation.} and STARLINK software packages. Initial reduction of the CCD frames involved bias substraction, flat-field correction and cosmic rays removal. In order to obtain the nuclear spectra, central apertures on the objects have been extracted. Wavelength calibration has been performed using ThAr lamp exposures. The wavelength calibration accuracy reached was always better than 0.5 and 1~\AA~in the residuals for NGC~3599 and SDSS~J1323 respectively.

From the flux calibrated nuclear spectrum of NGC~3599, we obtain [N~II]~6584~\AA/H$\alpha$=1.1($\pm$ 0.1). This emission line ratio stays constant with respect to the pre-outburst observation. The [O~III]~5007~\AA/H$\beta$ ratio differs from that of the pre-flare exposure. This is due to calibration issues in the optical data and not from real variability of the lines as could be stated from a posterior observation of the object (see below). The current spectrum does not exhibit any dramatic emission-line response, such as prominent broad Balmer lines or strong high-ionization lines as it has been recently discovered in the tidal disruption candidate SDSS~J095209.56+214313.3 \citep{Kom08}. The post-outburst spectrum of SDSS~J1323 showed no emission lines nor apparent variation since the spectrum performed before the disruption event. Due to the faintness of the source and the relatively short exposure time of the observation, a reliable flux calibration could not be achieved. However, this is not required for our aims as it has been confirmed that the source does not show at present any evident optical signature of the flaring event.

A new optical observation of NGC~3599 was performed in April 2008 with the IDS spectrograph on the 2.5~m Isaac Newton Telescope (INT) in La Palma. Given the strong underlying host galaxy absorption features, uncertainties in existing spectra are too large to assess small-amplitude variability of the emission lines. This observation was aimed to obtain for the first time a high quality spectrum of the source. Three exposures of 1800~s of the object were performed using the R300V grating and then analysed following the same procedure outlined for the NOT observations. A S/N$\sim$90 was achieved in the final spectrum. The linear fit applied between pixel number and wavelength for the selected spectral lines gave a calibration error less than 0.2~\AA. A set of 14 star exposures were observed and analysed in order to optimise the flux calibration process. Sensitivity functions of all standards were derived and combined into a unique solution using an iterative process that rejected those observations whose sensitivity behaviours deviated more than two percent from the mean calibration function. Gaussian profiles were fitted to the emission lines of the resulting nuclear spectrum of NGC~3599 after the starlight component subtraction (Fig.~\ref{fig:NGC3599_INT}). The following ratios were obtained: [O~III]~5007~\AA/H$\beta$=3.38($\pm$ 0.91), [N~II]~6584~\AA/H$\alpha$=0.99($\pm$ 0.07), [S~II]~6717+6734~\AA/H$\alpha$=0.31($\pm$ 0.05), [O~I]~6300~\AA/H$\alpha$=0.11($\pm$ 0.03). The source is located in the transition region between Seyfert and LINERS \citep{Kauffmann} and, presenting a ${L}_{{\rm H}_{\alpha}}$=10$^{39}$ erg s$^{-1}$, can be considered as a low luminosity AGN \citep{Ho1997}.

\section{X-ray light curves of the flare events}\label{sec:lightcurve}
The X-ray flare begins when the most bound material returns to pericentre after completing one orbit. Assuming that the mass distribution of the debris is nearly uniformly distributed as derived from numerical simulations \citep{Evans, Ayal}, the mass rate after one post-disruption orbit is

\begin{equation}
        \dot{M} \approx  \left(\frac{t - t_{\rm disr}}{1 {\rm yr}}\right)^{-5/3} \,
        \label{eq:mass_rate}
\end{equation}

\noindent
where $t_{disr}$ is the time at which the disruption process occurred.

\begin{figure*}[ht]

\begin{tabular}{cc}
\hspace{-1.6cm}
{\rotatebox[]{90}{\includegraphics[width=8.5cm,clip=true]{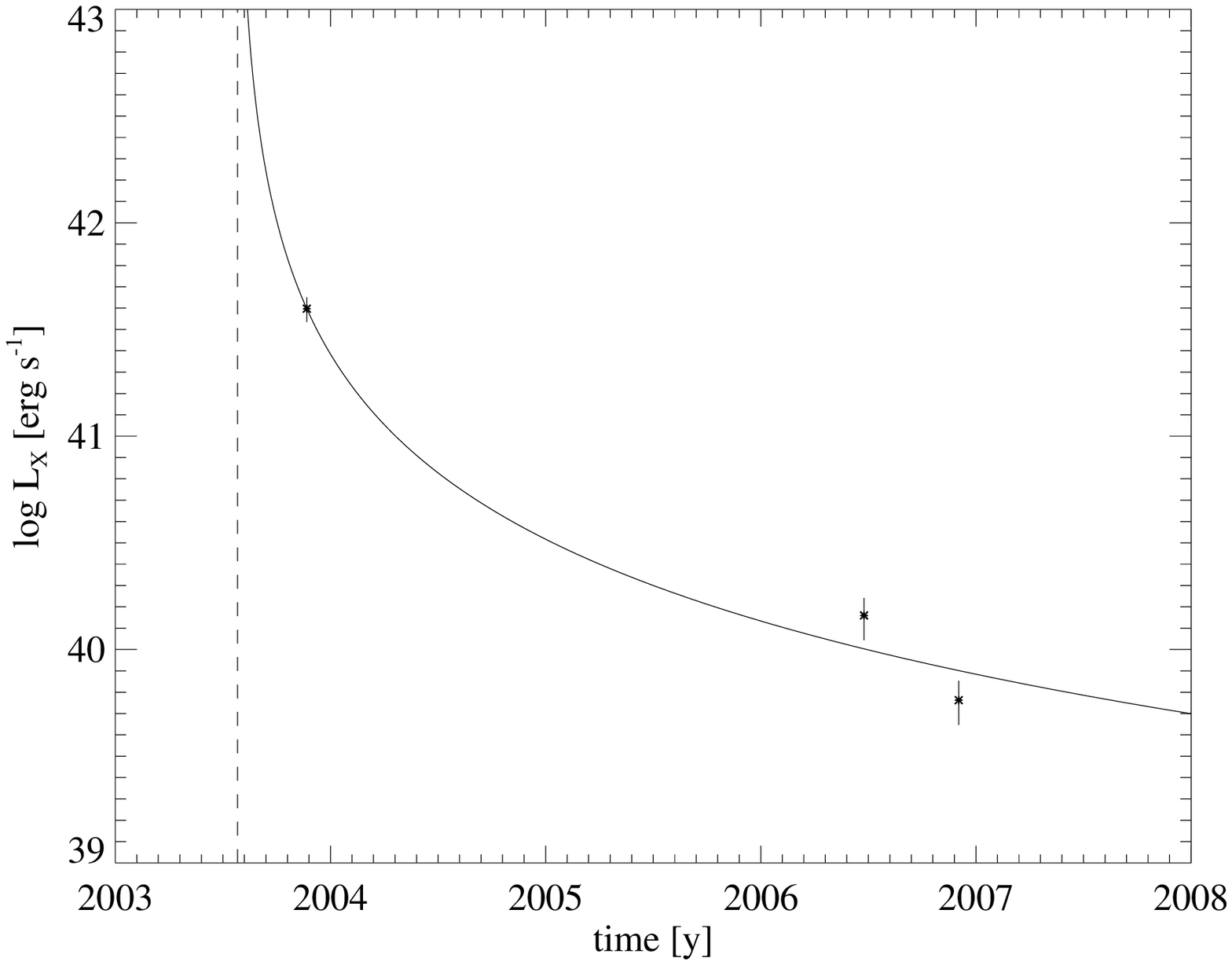}}}
\hspace{-2.5cm}
{\rotatebox[]{90}{\includegraphics[width=8.5cm, clip=true]{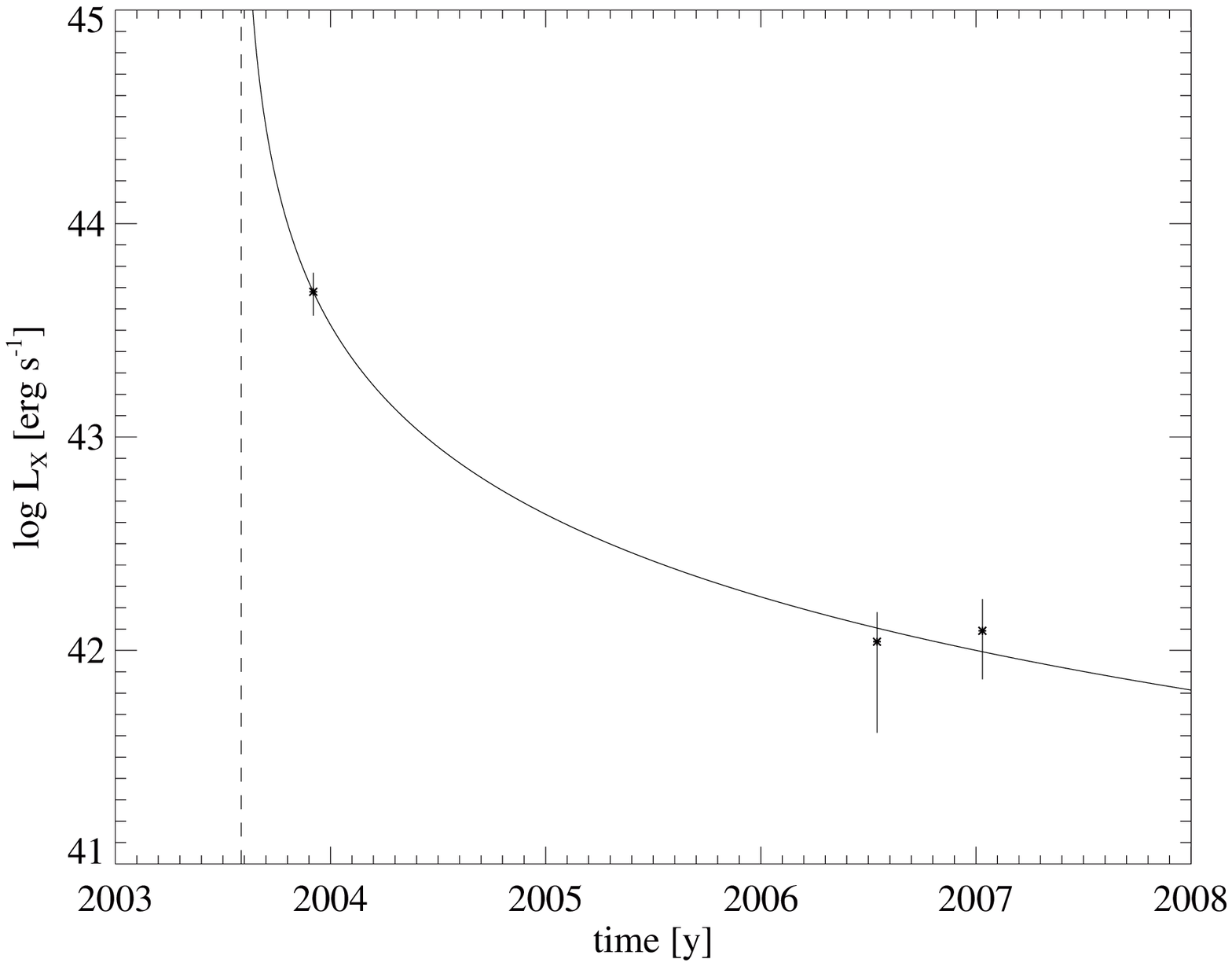}}}
\end{tabular}
\vspace{-2.5cm}
     \caption{X-ray light curve of NGC~3599 (left panel) and SDSS~J132341.97+482701.3 (right panel). In each plot, data points (90\% confident errors are included) correspond to XMM-slew, XMM-pointed and \emph{Swift} observations from left to right. Solid lines are the fitted curves predicted for the fallback stage and follow eqs. (\ref{eq:lum_X1}) and (\ref{eq:lum_X2}) respectively for both sources. Dashed lines show the disruption time (${t}_{\rm disr}$) as returned by the fitting.
               }  
        \label{fig:lightcurve}
\end{figure*}

Luminosities in the soft 0.2-2.0 keV energy band have been derived from all three X-ray source detections. Assuming that the debris evolution of our sources is represented by the fallback model of tidal disruption, data points in the light curve have been fitted with a $(t-t_{\rm disr})^{-5/3}$ decline law. These fits give

\begin{equation}
        L_X = {6.7} {(\pm~1.2)} \times 10^{40}\, 
              \left[\frac{t - {2003.59} {(\pm~0.04)}\, {\rm yr}}{1\,
              {\rm yr}}\right]^{-5/3} \, {\rm erg\, s^{-1}}\,
        \label{eq:lum_X1}
\end{equation}

\noindent
and

\begin{equation}
        L_{\rm X} = {8.1} {(\pm~2.9)}\times 10^{42}\, 
              \left[\frac{t - {2003.56} {(\pm~0.10)}\, {\rm yr}}{1\,
              {\rm yr}}\right]^{-5/3} \, {\rm erg\, s^{-1}}\,
        \label{eq:lum_X2}
\end{equation}

\noindent
for NGC~3599 and SDSS~J1323 respectively. Fig.~\ref{fig:lightcurve} shows the evolution of the X-ray luminosity for both sources and corresponding fitted light curves according to the previous equations. Although for the slew detections the representative black body model (\emph{kT}=0.07~keV) for tidal disruption events has been used for the luminosity calculation, error bars for these data points comprise the values obtained when allowing a range of models for the outburst spectra (black body with temperature within 0.04--0.1 keV and power-law with X-ray slopes of 2.9--3.5).

In order to estimate the total X-ray energy released during the event, the previous luminosities have to be integrated over the outburst duration via

\begin{equation}
        \Delta {E}_{\rm X} = \int_{t}^\infty L_{\rm X} (t) {\rm d}t .
        \label{eq:energy}
\end{equation}

\noindent
The flare, that begins when the most bound material returns to the central black hole after completing one orbit, occurred between the disruption time previously calculated (see eqs.~\ref{eq:lum_X1} and \ref{eq:lum_X2}) and the slew detection. We will use the time of the slew observations as conservative assumption for the beginning of the flare (lower limit of the integral in eq.~\ref{eq:energy}) in the following calculations. With this hypothesis, the total energy released during the entire event has been estimated for both objects. The associated total masses involved in the process have been derived by applying

\begin{equation}
        \Delta M = \frac{\Delta E}{\epsilon {c}^2} \approx
                   \frac{\Delta {E}_{\rm X}}{\epsilon {c}^2} 
        \label{eq:delta_mass}
\end{equation}

\noindent
where $\epsilon$ is the efficiency of converting mass to radiated energy (we used a value of $\epsilon$=0.1). It has been assumed that the bolometric correction is not very significant, which is a fair approximation if the spectrum is well represented by a black body as we expect near outburst. Assuming a black body model at 70 eV at the corresponding redshift of our sources, the bolometric correction would imply less than a factor of two. These results are presented in Table~\ref{tab:numbers} together with the black body radius of the emitting region \citep[as derived in][]{Li} expressed by

\begin{equation}
        R_{\rm X} = \left(\frac{f_{\rm c}^4 L_{\rm X}}{\pi\sigma T_{\rm bb}^4}\right)^{1/2}
        \label{eq:radius}
\end{equation}

\noindent
where $f_{\rm c} \equiv {T_{\rm bb}}/{T_{\rm eff}}$ and has been set to $f_{\rm c}$=3 for our calculations. Although the luminosity behaviour is not well constrained in the vicinity of the flare onset, we can also perform our calculations assuming that the beginning of the flare occurs at a time after the $\Delta{t}$\,$\approx$\,0.021 \citep[][using a 10$^6$ $M_{\odot}$ black hole and a solar-type star]{Li}, after the disruption. In that case, the estimated parameters in Table~\ref{tab:numbers} would increase by about an order of magnitude.

\begin{table}[ht]

\caption{Estimated properties of the disruption events. Column 1: source name. Column 2: energy released during the observed outburst. Column 3: accreted mass associated with the total event. Column 4: black body radius of the emitting region.}
\label{tab:numbers}
\centering

\sbox{\strutbox}{\rule{0pt}{0pt}}
\begin{tabular}{c|c|c|c}
\hline
& & & \\[\smallskipamount]
\textbf{Source} & \textbf{$\Delta {E}_{\rm X}$} (erg) & \textbf{$\Delta M$} ($M_{\odot}$) & \textbf{${R}_{\rm X}$} (cm) \\ 
& & & \\[\smallskipamount]

\hline
& & & \\[\smallskipamount]
NGC~3599 & \tiny{$7.1(\pm 0.8)\times10^{48}$} & \tiny{$4.0(\pm 0.4)\times10^{-5}$}  &  \tiny{$7.3(\pm 2.7)\times10^{11}$}  \\
& & & \\[\smallskipamount]

\hline
& & & \\[\smallskipamount] 

SDSS J1323 & \tiny{$7.6(\pm 0.3)\times10^{50}$} & \tiny{$4.2(\pm 0.2)\times10^{-3}$}  &  \tiny{$6.8(\pm 3.4)\times10^{12}$}  \\
& & & \\[\smallskipamount]
\hline

\end{tabular}
\end{table}

Several scenarios can be argued in order to explain the low amount of accreted matter. First, it may imply that we have only seen a partial disruption event. Conversely, it can also be indicative of a strong encounter when the star deeply penetrates within the tidal radius leading to a tidal detonation event. In that latter case, the star would experience a strong pre-disruption compression triggering thermonuclear reactions and being followed by the explosion of the star \citep[e.g.][]{Carter}. Therefore, only a small fraction of the debris would be actually accreted and the rest of the material being expelled \citep{Brassart}. The tidal radius (for a Sun-like star) as given by eq.~(\ref{eq:tidal_radius}) and the Schwarzschild radius are

\begin{equation}
R_{\rm T} \sim 7 \times 10^{12} \left(\frac{M_{\rm BH}}{10^6 M_{\odot}}\right)^{1/3} {\rm cm},
\hspace{0.7cm}
r_{\rm S} \sim 3 \times 10^{11} \left(\frac{M_{\rm BH}}{10^6 M_{\odot}}\right) \hspace{0.2cm} {\rm cm}.
\label{eq:radii}
\end{equation}

\noindent
The possibility of a stong encounter would be reinforced, especially for NGC~3599, as the derived radius of the emitting region lies inside the tidal radius for a 10$^6$ $M_{\odot}$ black hole. Nevertheless, further interpretation is out of scope as these values are simple order of magnitude estimates.

It is demonstrated that X-ray and optical luminosities of early-type galaxies present a tight correlation \citep{OSullivan}. Studies on NGC~5905, a previous case of tidal disruption, showed that the optical luminosity of the source did not experience any significant variation due to the flaring event \citep{KomBade}. We will assume that this is also applicable for our sources and we will use pre-outburst photometric data for our calculations (no relevant low-state information is available for comparison). B band magnitudes were obtained from the NED database (for SDSS~J1323 conversion factors have been applied to the Gunn magnitudes to derive the magnitude in the B Johnson filter). At the time of the \emph{Swift} observation we derive ratios (${log~L}_{X}/{L}_{B}$)=30 and 32 (with ${L}_{B}$ in units of solar luminosity) for NGC~3599 and SDSS~J1323 respectively. These values are beyond the observed limiting value for early-type galaxies and, therefore, we can argue that the X-ray emission of the sources is most likely related to the flare event.

An alternative estimation to characterise the nature of the low state emission of our sources can be reported by using the correlation of [OIII] and hard X-ray luminosities ($L_{\rm 2-10~keV}$ vs $L_{\rm [OIII]}$) of Seyferts and LLAGN. $L_{\rm [OIII]}$ luminosities have been used to predict the expected hard X-ray luminosities following eq. (1) in the recent estimation of \citet{Netzer} from their X-ray selected sample of sources. These values have been then compared with the X-ray luminosities in the 2--10 keV energy band obtained by extrapolating the best fit power-law model in the \emph{XMM--Newton} pointed observations. As can be seen from Fig.~\ref{fig:XMM_spectra}, sources are not very well modelled in this energy range and data points lie above the model, implying that our estimation falls below the real value of the luminosity. No significant change in spectral shape due to the flaring events is present in the optical data of both sources, indicating that the we could use either the pre- or post-outburst value of the [OIII] emission. For our calculation we used the [OIII] value from the highest resolution available spectra for both sources, i.e. the most recent optical observation of NGC~3599 and the pre-outburst observation for SDSS~J1323. For NGC~3599, the estimated 2-10~keV X-ray luminosity through the [OIII] correlation is $6\times 10^{38}$ erg~s$^{-1}$. This value is consistent (within the errors) with the $1.5\times 10^{39}$ erg s$^{-1}$ luminosity returned by the spectral fit from section~\ref{subsubsec:XMM_obs}. Although we can not constrain with this argument that the low-state X-ray emission is not AGN related, it is worth mentioning that the equation used for predicting the X-ray luminosity was derived for luminosities down to L$_{\rm 2-10 keV}$=$10^{41}$ erg~s$^{-1}$. As NGC~3599 presents a luminosity two orders of magnitude lower, this can conflict with the validity of the formula for our calculation. The optical spectrum of SDSS~1323 does not present emission line features. A single gaussian fit has been performed to the position of the [OIII] emission line and a 3$\sigma$ upper limit on the $L_{\rm [OIII]}$  has been derived, with redshift and linewidth fixed to that determined from stellar absorption lines. In this case, the $L_{\rm 2-10~keV}$=$3.4\times 10^{39}$ erg~s$^{-1}$ obtained from the relationship in \citet{Netzer} is below the $4.3\times 10^{40}$ erg~s$^{-1}$ obtained from the spectral fit. As the low-state X-ray luminosity for SDSS~1323 is higher than that predicted from the [OIII] emission, we have an extra argument to confirm that the low-state emission is flare related, rather that coming from a permanent AGN.

\section{Black hole mass estimates}\label{sec:bh_mass}
Due to evidence for the existence of a massive black hole located at the centres of most galaxies, a number of empirical relationships have been derived relating the black hole mass and different characteristics of the corresponding host galaxies. 

Among them is the $M_{\rm BH}-\sigma$ relationship, that is the correlation between the black hole mass and the stellar velocity dispersion. Using the recent estimation from \citet{Ferrarese2005} 
\begin{equation}
M_{\rm BH} = 1.66 (\pm 0.24) \times 10^8 M_{\odot} \left({\frac{\sigma}{200~{\rm km}~{\rm s}^{-1}}} \right)^{4.86(\pm 0.43)}
\label{eq:Mass_sigma_rel}
\end{equation}
\noindent
we derive a black hole mass for NGC~3599 of 1.3($\pm 0.6$)$\times10^{6} M_{\odot}$ using $\sigma$=73.2 km s$^{-1}$ (several velocity dispersion measurements exist for this galaxy and the median value provided by Hyperleda as been used in our calculation). For SDSS~J1323 the mass is 2.2($\pm 0.9$)$\times10^{6} M_{\odot}$, with $\sigma$=80 km s$^{-1}$ as obtained from the analysis of the optical spectrum (see section \ref{subsec:J1323}) (no other measurement is available for comparison). 

Conversely, using the $M_{\rm BH}-\sigma$ estimate from \citet{Lauer2007}

\begin{equation}
M_{\rm BH} = 1.95 (\pm 1.17) \times 10^8 M_{\odot} \left({\frac{\sigma}{200~{\rm km~s}^{-1}}} \right)^{4.13(\pm 0.32)}
\label{eq:Mass_sigma_rel_2}
\end{equation}

\noindent
we obtain 3.1($\pm 1.0$)$\times10^{6} M_{\odot}$ and 5.0($\pm 1.4$)$\times10^{6} M_{\odot}$ for NGC~3599 and SDSS~J1323 respectively.

Other estimation arises from the relationship between the black hole mass and the V absolute magnitude of the host galaxy. Using the correlation derived in eq. (6) of \citet{Lauer2007b}, black hole masses of 3.5($\pm 0.9$)$\times10^{7} M_{\odot}$ and 3.3($\pm 0.9$)$\times10^{7} M_{\odot}$ are obtained for NGC~3599 and SDSS~J1323 respectively. 

Black hole mass estimates from both $M_{\rm BH}-\sigma$ and $M_{\rm BH}-{L}_{\rm V}$ approaches differ by one order of magnitude approximately, due to uncertainties in the measurements and scatter in the correlations. Nevertheless, results from both calculations lie within the mass range expected for objects experiencing a tidal disruption.

Using the black hole masses derived from the different approaches, the observed high-state luminosity results in a fraction of the Eddington luminosity. That factor is up to 2\% for NGC~3599 and 17\% for SDSS~J1323 (using the smallest black hole mass estimation). Therefore, at the time of the slew observation the emission was sub-Eddington, especially for NGC~3599.

\section{Tidal disruption rate from the slew survey}\label{sec:tidal_rate}
The identification of tidal disruption events discussed here was enabled by the existence of two large area sensitive X-ray surveys performed at different epochs. We can only attempt to calculate a first order estimate of the outburst rate as a number of assumptions have to be taken into account.

In order to determine the characteristic outburst flux limit to which our survey is complete, we first derived the nominal RASS flux by computing $2\sigma$ upper limit count rates of 5000 sources randomly distributed over the whole sky, as we assumed that the outbursting galaxies are isotropically distributed. These rates in the 0.1--2.4 keV energy band were converted to 0.2--2 keV unabsorbed fluxes through PIMMS assuming the typical spectral model expected near the peak of the tidal event, i.e. a black body model with temperature 0.07 keV modified by galactic absorption\footnote[2]{Galactic absorption for each source position was calculated using the ${N}_{\rm H}$ tool provided by PIMMS.}. The flux distribution is widely spread due to the different exposure times of the RASS observations and the different absorption column densities of the sources. In order to avoid high galactic absorption and confusing sources in the galactic plane, in the remaining calculation we will not consider the region with Galactic latitude $\vert{\rm b}\vert < $ 20~deg. From the flux histogram, we derived a median flux of $2.5\times10^{-13}$~erg~${\rm s^{-1}}$~${\rm cm^{-2}}$. We will adopt this number as the characteristic unabsorbed flux limit for the RASS, which is in fair agreement with the $3\times10^{-13}$~erg~${\rm s^{-1}}$~${\rm cm^{-2}}$ typically quoted \citep{Brandt}. Therefore, if we consider as outbursting events those sources which varied by at least a factor of 20 with respect to the RASS, the characteristic flux to which our survey is complete is $5.08\times10^{-12}$~erg~${\rm s^{-1}}$~${\rm cm^{-2}}$. The on-axis flux limit in the soft band for slew observations applying the black body model is $4.33\times10^{-13}$~erg~${\rm s^{-1}}$~${\rm cm^{-2}}$ and raises to $4.33\times10^{-12}$~erg~${\rm s^{-1}}$~${\rm cm^{-2}}$ for the minimum exposure times, hence we reach the required sensitivity for completion derived from the RASS picture in the whole slew coverage. Tidal disruption events near maximum are expected to emit a luminosity up to $10^{44}$ $-$ $10^{45}$~erg~${\rm s^{-1}}$ depending on the black hole mass. Using an outburst luminosity of $10^{44}$~erg~${\rm s^{-1}}$ (the median value for our two candidates) and the flux limit above calculated we derived that the completeness distance of our survey is 406 Mpc (z=0.089) that, in terms of volume, corresponds to $2.8\times10^{8}$ ${\rm Mpc}^{3}$.

\begin{figure}
\centering
   {\rotatebox[]{90}{\includegraphics[width=7cm]{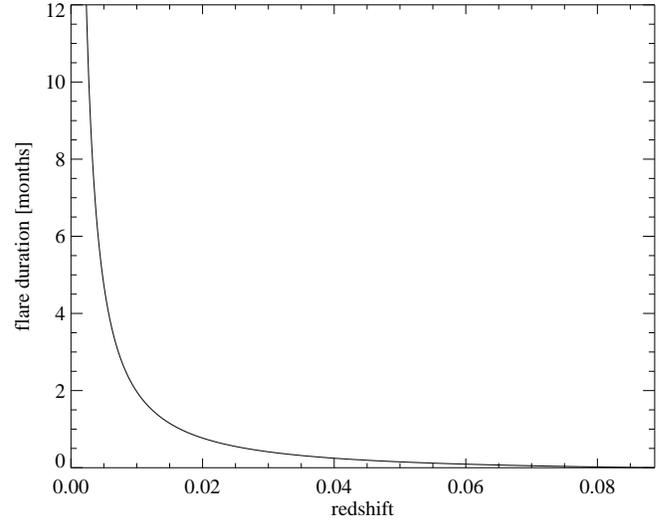}}}
\vspace{-0.7cm}    
\caption{Observable flare duration depending on redshift.}
\label{fig:flare_redshift}
\end{figure}

In order to derive the tidal disruption rate we will have to consider the time interval in which the flare, that starts when the most bound material returns to pericentre, is visible before the source has declined to non-detectability. To estimate this figure, we will assume that for nearby sources the galaxy luminosity peaks at $10^{44}$~erg~${\rm s^{-1}}$, follows the $t^{-5/3}$ law and then declines to its quiescent value in three years after outburst. Whatever the morphologycal type, the luminosity of a normal galaxy lies in the range $10^{38}$--$10^{42}$~erg~${\rm s^{-1}}$ \citep{Fabbiano}, so we presupposed a dormant luminosity of $10^{40}$~erg~${\rm s^{-1}}$. The next step will be to simulate that curve including the redshift dependence by means of the flux decrease due to the geometric dilution factor, as the unabsorbed X-ray luminosity depends on the squared luminosity distance (${L}_{\rm X}=~{F}_{\rm X}4\pi{D}_{\rm L}^2$). With these hypotheses, all of them in fair agreement with theory, we derive that for sources up to redshift 0.0025 we will see the events at least for one year before declining below our limiting flux of $5.08\times10^{-12}$~erg~${\rm s^{-1}}$~${\rm cm^{-2}}$. The observable duration of the flare depending on redshift and estimated under these premises is presented in Fig.~\ref{fig:flare_redshift}. 

To obtain the number of events we detect per year per unit volume we will divide the number of tidal candidates by

\begin{equation}
{\int_{0}^{R_{\rm max}} A(r)t(r){\rm d}r}
\end{equation}

\noindent
where $R_{\rm max}$=406 Mpc as derived before, \emph{A(r)} represents the area covered (including the 16.5\% fraction of the sky surveyed by slew obsevations in the XMMSL1\_Delta1\footnote[3]{http://xmm.esac.esa.int/external} with $\vert{\rm b}\vert < $ 20~deg), and \emph{t(r)} is the observable flare duration as observed in Fig.~\ref{fig:flare_redshift}. As we detected two tidal diruption candidates, the outburst rate we obtain is 5.4$\times10^{-6}$ ${\rm yr}^{-1}$ ${\rm Mpc}^{-3}$.

In order to estimate the rate of events per galaxy we need to take into account the total galaxy space density. Extensive work has been done in deriving luminosity functions of galaxies depending on their morphology, as different galaxy types require a distinct modelling using either Gaussian or Schechter functions. That topic is beyond the scope of the present paper and we will simply use the derived parameters from \cite{deLapparent}  for our calculation. In the same way as in other attempts of calculating the tidal disruption rate, i.e. including all E, S0 and elliptical galaxies in the B band for the space density calculation, we derived a total space density of $2.3\times10^{-2}$ ${\rm Mpc}^{-3}$. Therefore, the resulting tidal disruption rate is $2.3\times10^{-4}$ ${\rm galaxy}^{-1}$ ${\rm yr}^{-1}$.

In order to validate our result, we have compared our tidal disruption rate with that derived from theoretical predictions. The tidal disruption rate as estimated in \citet{Wang} can be expressed by

\begin{equation}
\dot{N}\approx 7.1\times 10^{-4} {\rm yr}^{-1}
\left({\frac{\sigma}{70\ {\rm km\ s}^{-1}}}\right)^{7/2}
\left(\frac{M_{\rm BH}}{10^6 M_{\odot}}\right)^{-1}
\left(\frac{m_\star}{M_\odot}\right)^{-1/3}
\left(\frac{R_\star}{R_\odot}\right)^{-1/4}
\label{eq:tidal_rate_Wang}
\end{equation}

\noindent
$\sigma$ being the velocity dispersion of the host galaxy, and $m_\star$ and $R_\star$ the corresponding mass and radius of the disrupted star. This equation was normalised to reproduce the tidal disruption rate for $\sigma$=70~km~s$^{-1}$ and a $10^{6}$~$M_{\odot}$ black hole. The dependence on the stellar mass and radius is weak as the tidal disruption rate is expected to be dominated by subsolar stars (which fulfill the relation $R_\star$/$R_\odot \approx m_\star$/$M_\odot$). We will assume solar mass and radius for simplicity. Merging eqs. (\ref{eq:Mass_sigma_rel}) and (\ref{eq:tidal_rate_Wang}), the unique dependence of the tidal rate on the black hole mass becomes explicit. Considering a $10^{6}$~$M_{\odot}$ black hole, the consumption rate is 7.0$\times10^{-4}$ ${\rm galaxy}^{-1}$ ${\rm yr}^{-1}$. Note that eq.~(\ref{eq:tidal_rate_Wang}) only provides a good parameterisation for power-law galaxies with $\Gamma > $~0.9, overestimating the tidal disruption rate by up to an order of magnitude for galaxies with shallower power-law and core brightness profiles. Although a number of simplifications have been included in our estimations, our predicted tidal disruption rate as derived from the observations agrees with that from the theoretical approach.

\begin{figure}
\centering
\vspace{-0.7cm}
   {\rotatebox[]{90}{\includegraphics[width=7.75 cm, height=9.8cm]{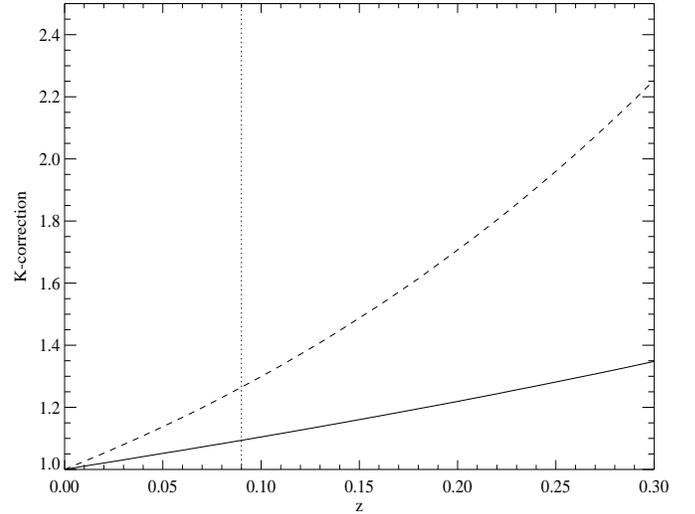}}}
\vspace{-1.3cm}    
\caption{K-correction. Solid line: \emph{XMM--Newton} EPIC-pn. Dashed line: \emph{ROSAT} PSPC. Dotted line: limiting redshift for completeness in our calculation.}
\label{fig:kcorrection}
\end{figure}

For completeness, although not very important as we are dealing with nearby objects, we have estimated the dimming effect on the flux determination due to the redshifted energy band, i.e. the \emph{K}-correction, that in our case can be expressed by

\begin{equation}
{K}_{0.2-2.0~{\rm keV}}= 
\frac{\int_{0.2}^{2.0} f_{\nu} {\rm d} \nu} 
{\int_{0.2(1+z)}^{2.0(1+z)} f_{\nu} {\rm d} \nu}
\end{equation}

\noindent
where $f_{\nu}$ is the unabsorbed spectral flux density. X-ray spectra have been simulated in XSPEC to derive that bandpass correction for a redshifted black body model (zblackbody) at \emph{kT}=70~eV using both EPIC-pn and \emph{ROSAT} response files. The galactic absorption has been taken as ${N}_{\rm H}$=$4.03\times10^{20}$~cm$^{-2}$, which is the median value for the random positions in the sky calculated at the beginning of this section. This effect, that is significantly smaller than the geometric dilution factor, produces a decrease up to 10\% for the \emph{XMM} and 27\% for the ROSAT fluxes at our redshift limit (Fig.~\ref{fig:kcorrection}).

We have performed our calculations presupposing that intrinsic absorption does not prevent the detection of the flaring events. As discused in \citet{Sembay}, this effect can by included by estimating the rate of non edge-on spiral galaxies in our sample. Due to the large uncertainty existent in the derivation of that number, the former calculation did not include this argument. As a rough approximation, assuming a fraction of 60\% for objects in which intrinsic absorption does not play an important role \citep{Donley}, the tidal disruption rate would increase up to $3.9\times10^{-4}$ ${\rm yr}^{-1}$ ${\rm Mpc}^{-3}$.

\section{Summary and conclusions}
We have presented X-ray and optical follow-up observations of two tidal disruption candidates detected by \emph{XMM--Newton} during slew observations. 

The softness of our sources near outburst confirms that tidal disruption events are certainly very soft. Both NGC~3599 and SDSS~J1323 were not detected in the hard 2--12 KeV slew energy band close to the peak of the outburst.  As all previous X-ray candidates were detected with \emph{ROSAT}, the spectral softness near outburst could have been biased due to selection effects. However, this is unlikely because they were very soft even within the \textit{ROSAT} 0.1--2.4 keV spectral coverage.

Although tidal disruption phenomenon is the most plausible scenario in order to describe the events here discussed, extreme variability of a possible ultraluminous X-ray source (ULX) located within NGC~3599 can not be safely excluded with the present data due to the relatively low luminosity of the source near outburst. Nevertheless, a ULX interpretation appears to be \textit{a priori} improbable as this class of objects are characterised by X-ray luminosities in the range $10^{39}$--$10^{40}$ erg~s$^{-1}$, flux variability factors of 2--3 and a power-law spectral shape with X-ray photon index between 1.6 and 1.8 \citep{Colbert2002}.

\emph{XMM--Newton} and \emph{Swift} observations of the two tidal disruption candidates described in this paper have been performed in order to monitor the evolution of the objects. These are the nearest hard ($>$~2 keV) X-ray follow-up observations to an X-ray high state detection ever performed. We can now assess the existence of hard X-ray photons within three years after the observed high state, although a spectral hardening can not be quantified as this is highly dependent of the spectrum assumed near outburst. Prior to this work, a single reliable hard band modelling of the phenomena was only achieved ten years after the maximum \citep{Kom04}. No absorption excess to the Galactic value is required in order to fit the spectra from the present data, although there is some evidence for an edge in the SDSS~J1323 spectrum. With power-law slopes of $\Gamma_{\rm X}$=3 and 3.5, the spectra are still very soft compared with normal AGN. We tentatively interpret the hard emission as coming from an accretion disk although in the case of NGC~3599 emission from the galaxy itself cannot be excluded. 

The X-ray flux is seen to decrease between the \emph{XMM--Newton}-slew and \emph{Swift} follow-up observations by factors of 30 and 40 respectively for NGC~3599 and SDSS~J1323. The light curves plausibly follow the expected \textit{t}$^{-5/3}$ curve. Post-flare optical spectra of both sources did not reveal any dramatic emission-line response due to the flaring event. 

After detailed optical spectroscopic analysis, NGC~3599 turned out to be misclassified as a inactive galaxy and appears to harbor a dwarf active nucleus. This fact is not unexpected considering that the majority of the nearby galaxies host weakly active AGN-like cores \citep{Ho2001}. This interpretatation was also inferred for the previous tidal disruption candidate NGC~5905 due to the detection of faint \emph{HST}-based post-flare high ionization lines \citep{Gezari}. Nevertheless, given that there is no \emph{HST} spectrum of NGC~5905 prior to the outburst, the emission lines detected in this source could also have been excited by the flare itself as seen in a very recent candidate \citep{Kom08}.

The tidal disruption rate as derived from slew observations is $2.3\times10^{-4}$ ${\rm galaxy}^{-1}$ ${\rm yr}^{-1}$, which agrees with previous theoretical estimations, and further endorses the view that these X-ray flares are the result of the tidal disruption by a SMBH.

The detection of these phenomena represents a unique tool to unveil the dormant black hole population at the centres of the flaring galaxies. A complete census of the relic SMBHs is fundamental to constrain the accretion history of the Universe. It has been suggested that tidal disruption events could fuel the majority of low luminosity AGN (LLAGN) and perhaps significantly contribute to their luminosity function \citep{Milosavljevic}. More flare events are expected to emerge in the near future based on the data collected during \emph{XMM--Newton} slew observations besides future X-ray surveys like that planned with eROSITA.

\begin{acknowledgements}
We would like to thank Nelson Caldwell for providing the optical data used in section \ref{subsec:NGC3599}. The \emph{XMM--Newton} project is an ESA Science Mission with instruments and contributions directly funded by ESA Member States and the USA (NASA). The XMM-Newton project is supported by the Bundesministerium f\"ur Wirtschaft und Technologie/Deutsches Zentrum f\"ur Luft- und Raumfahrt and the Max-Planck Society. Part of this work is based on observations made with the NOT and INT Telescopes under the Spanish Instituto de Astrof\'{i}sica de Canarias CAT Service Time. Data presented here have been taken using ALFOSC, which is owned by the Instituto de Astrof\'{i}sica de Andaluc\'{i}a (IAA) and operated at the Nordic Optical Telescope under agreement between IAA and the NBIfAFG of the Astronomical Observatory of Copenhagen. The INT is operated on the island of La Palma by the Isaac Newton Group in the Spanish Observatorio del Roque de los Muchachos of the Instituto de Astrof\'{i}sica de Canarias. PE acknowledges support from the International Max Planck Research School on Astrophysics and the European Commision under a Marie Curie Host Fellowship for Early Stage Researchers Training, and AMR the funding support of PPARC/STFC. 
\end{acknowledgements}

\bibliographystyle{aa}
\bibliography{bibliography}

\end{document}